\documentclass{IEEEtran}
\usepackage{cite}
\usepackage{amsmath,amssymb,amsfonts}
\usepackage{algorithmic}
\usepackage{graphicx}
\usepackage{textcomp}
\usepackage{enumitem}

\usepackage{mathtools}
\def\BibTeX{{\rm B\kern-.05em{\sc i\kern-.025em b}\kern-.08em
    T\kern-.1667em\lower.7ex\hbox{E}\kern-.125emX}}
\begin{document}
\title{Low-profile Aperiodic Metasurfaces for High-efficiency Achromatic Anomalous Reflection over a Wide Bandwidth}
\author{Vasileios~G.~Ataloglou, \IEEEmembership{Member, IEEE}, and George~V.~Eleftheriades, \IEEEmembership{Fellow, IEEE}
\thanks{V.~G.~Ataloglou and G.~V.~Eleftheriades are with the Edward S. Rogers Sr. Department of Electrical \& Computer Engineering, University of Toronto, Toronto ON M5S3G4, Canada (e-mail: vasilis.ataloglou@mail.utoronto.ca, gelefth@waves.utoronto.ca).}}
\maketitle
\begin{abstract}
Metasurface anomalous reflectors can enhance future wireless communication systems due to their ability to redirect electromagnetic waves and establish metasurface-assisted communication links. However, metasurface reflectors typically suffer from chromatic aberrations with the reflected angle shifting as the frequency varies. Herein, we design achromatic metasurfaces for anomalous reflection by using an integral-equation framework that fully accounts for the frequency dispersion of the individual scatterers (unit cells). Rather than engineering the dispersion of each metasurface cell locally, we harness the emerging near-field interactions (evanescent waves) between the scatterers to realize anomalous reflection at a fixed angle over a wide frequency range. Three anomalous reflectors are designed and experimentally verified at 10GHz, demonstrating high performance metrics compared to traditional achromatic reflectors, while employing simple, low-profile unit cells. The presented results pave an effective way to realize achromatic wave functionalities, such as anomalous reflection, refraction or lensing, over a wide bandwidth with low-profile and highly efficient metasurfaces.
\end{abstract}


\section{Introduction}
\label{sec:introduction}
Metasurfaces have demonstrated remarkable potential to efficiently control electromagnetic waves and realize various functionalities, such as anomalous refraction or reflection, beam focusing and polarization control \cite{Yu:Science2011,Pfeiffer:PRL2013,Selvanayagam:Opt2013,Estakhri:PRX2016,Lavigne:TAP2018,Chen:TAP2019WideAngle,Wong:PRX2018,Chen:TAP2019Lens,Xue:TAP2020,Niemi:TAP2013}. Recently, the integration of metasurfaces in wireless communication systems has been proposed in order to establish a smart radio environment, where reflective metasurfaces are utilized in order to establish communication links in areas not sufficiently covered by the existing base stations \cite{DiRenzo:JSAC2020,Salucci:TAP2023}. Both static and reconfigurable metasurfaces have been developed to this purpose, enabling the reflection of one or more incident beams to desired reflected directions \cite{Benoni:TAP2023,Keyrouz:APL2023,Sievenpiper:TAP2003,Cui:TAP2003,Yang:TAP2016,Araghi:Access2022}. However, the strong resonances of the scatterers comprising the metasurface and the usually periodic nature of the entire structure typically narrow the operating bandwidth and lead to chromatic aberrations that limit the usefulness of metasurfaces in  wireless communications. In particular, it is vital to design metasurfaces that can efficiently reflect a wave from a fixed incident angle $\theta_i$ to a fixed reflected angle $\theta_r^0$ over a wide bandwidth. The realization of such achromatic anomalous reflectors would allow multiple spectrally-multiplexed channels to be reflected simultaneously through a single metasurface to a direction where an end terminal or a consecutive metasurface panel may be located. Additionally, these devices need to be low-profile and low-cost so that they can be easily incorporated into future generations of communications.

A phase-gradient metasurface for anomalous reflection would typically exhibit a reflected angle that varies with frequency, as dictated by diffraction optics \cite{Aieta:Sci2015,Arbabi:Opt2017}. On the contrary, one way to achieve an achromatic response is to engineer the dispersion of each individual scatterer comprising a planar metasurface. For anomalous reflection, the generalized law of reflection implies that the reflected phase of a scatterer (unit cell) at a position $x$ along the metasurface needs to be: $\Phi(x,f)= \Phi_0+(2\pi f x/c)[\mathrm{sin}(\theta_r^0)-\mathrm{sin}(\theta_i)]$ at each particular frequency $f$, with $\Phi_0$ being some arbitrary phase reference. Essentially, this requirement imposes not only a specific reflected phase for each unit cell, but also a specific group delay over the desired bandwidth making the design of a low-thickness sub-wavelength unit cell extremely challenging. Several achromatic metasurfaces have been designed, primarily in the mid-infrared to visible spectrum, for anomalous reflection, anomalous refraction and lensing \cite{Aieta:Sci2015,Arbabi:Opt2017,Khorasaninejad:NanoLet2015, Wang:NatCom2017, Shrestha:LightSci2018, Chen:NatNano2018, Chen:NatCom2019, Ou:SciAdv2020, Wang:AoM2022,Chen:NatCom2023}. Similar phase-compensation concepts have been applied at microwaves, where the use of copper-based cells is usually preferred compared to the use of a dielectric in order to achieve more compact designs \cite{Fathnan:AoM2020,Zhang:AoM2021,Yang:TAP2023,Ji:LPR2022,Yang:APL2016, Yang:APL2017,Jia:JoP2018,Zhu:OE2020,Yu:OME2022}. However, the existing achromatic metasurfaces for reflection or refraction rely on electromagnetically-thick structures (in the scale of half wavelength or more) to match the desired cell dispersion  \cite{Yang:APL2016, Yang:APL2017,Ji:LPR2022}, exhibit relatively low directivity \cite{Jia:JoP2018}, or they are constrained to small deflection angles and narrow bandwidths \cite{Ji:LPR2022,Zhu:OE2020,Yu:OME2022}.

The paper herein introduces the concept of a broadband achromatic reflector that is designed globally rather than on a cell-to-cell level. This approach relaxes the stringent requirement of engineering the cell dispersion and allows the realization of low-thickness unit cells at microwaves that collectively achieve the desired response. The underlying mechanism is fundamentally different and it relies on the excitation of surface waves at each frequency that supplement the incident and reflected waves in a way that both the dispersion model of the selected unit cell and passivity are satisfied. It is noted that auxiliary surface waves have been thoroughly investigated as a way to restore local power conservation at a single frequency by redistributing the incident power along the metasurface aperture \cite{Epstein:PRL2016,Kwon:PRB2018,Kwon:AWPL2021,Ataloglou:AWPL2021,Salucci:TAP2018}; yet, their utilization in achieving a broadband achromatic wave transformation remains unexplored. Moreover, the proposed metasurfaces are aperiodic to alleviate the limitations associated with the Floquet-Bloch (F-B) modes of periodic structures \cite{Li:TAP2024}. In particular, all periodic implementations for achieving highly-efficient anomalous reflection, such as periodic Huygens' metasurfaces exciting surface waves \cite{Epstein:PRL2016}, impenetrable periodic metasurfaces designed based on leaky-wave principles \cite{Diaz-Rubio:SciAdv2017}, or metagratings \cite{Radi:PRL2017,Wong:PRX2018,Rabinovich:PRB2019} are inherently affected by chromatic aberrations, as their scattering is expanded to a set of propagating F-B modes. Because the angle of each mode depends on the frequency, the reflected beam will not be angularly-stable, even if the metasurface is designed to reflect efficiently to a desired F-B mode over a wide bandwidth \cite{Qi:OE2022}.

The proposed method relies on a simplified model that treats the patterned unit cells as strips of homogenized surface impedance. Then, the metasurface is analyzed through a set of volume-surface integral equations that determine the reflected field, while taking into account all near-field interactions between the unit cells \cite{Budhu:TAP2020,Xu:Access2022}. The unit cell's surface impedance is extracted for different frequencies and geometries through full-wave simulations, and it is supplemented to the integral-equation framework in order to accurately predict the metasurface response over multiple frequencies. By optimizing the impedance values of each strip, an achromatic wave transformation can be realized, while the surface impedances range within the realizable limits and adhere to the cell's dispersion model. Specifically, for achromatic anomalous reflectors, the target is to maximize the directivity at a fixed reflected angle over a set of frequencies across the desired frequency range.

The rest of the paper is organized as follows. In Sec.~\ref{sec:optimization}, the framework to optimize a metasurface for realizing achromatic anomalous reflectors is described. The unit cell and the method to extract its homogenized impedance over multiple frequencies is also presented. Using the optimization framework, the efficiency of the anomalous reflection for various choices of the reflection angle and frequency bandwidth is investigated in Sec.~\ref{sec:simulation}. Additionally, the role of auxiliary surface waves in achieving the desired achromatic operation is analyzed. Section~\ref{sec:measurement} includes the experimental verification through the measurement of three achromatic reflector prototypes, while conclusions are drawn in Sec.~\ref{sec:conclusion}. 

\section{Optimization framework}\label{sec:optimization}

\subsection{Integral-equation formulation}
\begin{figure}
\centering
\includegraphics[width=0.95\columnwidth]{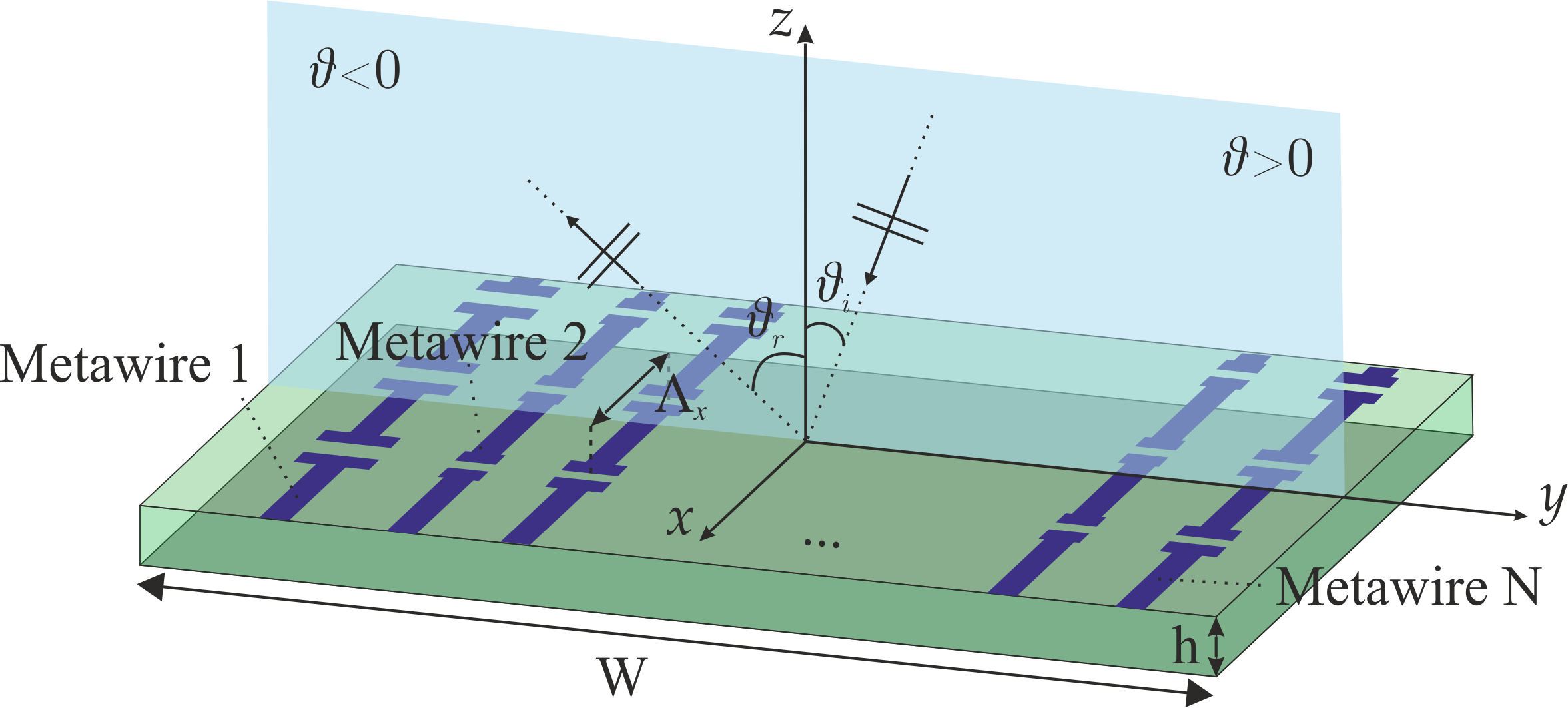}
\caption{\label{fig:Fig1} Sketch of the impedance MTS consisting of a set of $N$ columns of loaded meta-wires etched on a dielectric substrate. The aim is to design the impedance loadings such that anomalous reflection is achieved from an incident angle $\theta_i$ to a fixed angle $\theta_r^0$ over a wide bandwidth.}
\end{figure}

The design of the metasurface acting as an achromatic anomalous reflector is based on the accurate analysis of the structure through a set of volume-surface integral equations. The process has been analytically presented in \cite{Xu:Access2022} for the case of monochromatic illumination from an embedded source; yet, the basic steps are reviewed herein for completeness. Specifically, metasurfaces that vary along the $y$-direction while being uniform along the $x$-direction ($\partial/\partial x =0$) are considered, as sketched in Fig.~\ref{fig:Fig1}. The metasurface consists of an impedance layer of total length $L_\mathrm{tot}$ that lies on top of a grounded substrate of thickness $h$ and relative permittivity $\varepsilon_r$. The impedance layer is modelled with $N$ strips of width $w$ and frequency-dispersive homogenized impedance $Z_n(f)=R_n(f)+jX_n(f)$. In a second step, these impedances are realized by means of appropriately designed cells, as will become evident in the rest of the paper. For simplicity, we assume a transverse-electric (TE) incident plane wave $\left(\mathbf{E}^\mathrm{inc}=E^\mathrm{inc}(y,z) \mathbf{\hat{x}}\right)$ from an incident angle $\theta_i$ in the $yz$-plane. The purpose of the metasurface is to efficiently reflect the incident wave into a desired angle $\theta_r^0$ over a frequency range $[f_\mathrm{min},f_\mathrm{max}]$. By convention, we assume positive elevation angles ($\theta>0$) for the right-half of the $yz$-plane ($\phi=90^\circ$), and negative elevation angles ($\theta<0$) for the left-half of the $yz$-plane ($\phi=270^\circ$).

Upon incident illumination at a frequency $f$, a surface current density will be induced on the impedance sheets and the ground plane, denoted as $\mathbf{J}_w=J_w(y) \mathbf{\hat{x}}$ and $\mathbf{J}_g=J_g(y) \mathbf{\hat{x}}$, respectively. Similarly, a volumetric polarization current density $\mathbf{J}_v=J_v(y,z) \mathbf{\hat{x}}$ will be induced on the dielectric substrate. Since the structure extends infinitely along the $x$-axis, a cross-section can be analyzed through the free-space two-dimensional (2D) Green's function. At the $yz$ cross-section, the impedance strips collectively define a curve $C_w$, the ground plane defines a line segment $C_g$ and the substrate a surface area $S_v$. The scattered fields radiated from the induced currents are polarized along $x$, and they are calculated as:
\begin{align}
E_{i}^\mathrm{sc} (\mbox{\boldmath$\rho$})= \begin{dcases}
-\frac{k \eta}{4} \int_{C_i} J_i (y') H^{(2)}_0 (k|\mbox{\boldmath$\rho$}-\mbox{\boldmath$\rho'$}|) dl', \ i=\{w,g\}, \\
-\frac{k \eta}{4} \int_{S_i} J_i (y',z') H^{(2)}_0 (k|\mbox{\boldmath$\rho$}-\mbox{\boldmath$\rho'$}|) ds',  \  \  \  \ i=v,
\end{dcases}
\end{align}
where $i=\{w,g,v\}$ refers respectively to the impedance sheets, ground plane and dielectric substrate, $k=2\pi f/c_0$ is the free-space wavenumber ($c_0$ being the speed of light in vacuum), $\eta \approx 377 \Omega$ is the free-space impedance, $\mbox{\boldmath$\rho$}=y \mathbf{\hat{y}}+z \mathbf{\hat{z}}$ and $\mbox{\boldmath$\rho'$}=y' \mathbf{\hat{y}}+z' \mathbf{\hat{z}}$ are the position vectors of the observation point and the source location, respectively, and $H_0^{(2)}$ is the second-kind Hankel function of zeroth order. The total $x$-oriented electric field, consisting of the incident and scattered electric fields needs to be proportional to the induced currents on the impedance strips and in the dielectric, and it must vanish on the ground plane, according to:
\begin{align} \label{eq:VSIE}
E^\mathrm{inc}+E_w^\mathrm{sc}+E_g^\mathrm{sc}+E_v^\mathrm{sc}= \begin{dcases}
Z(y) J_w(y), & \mathrm{on} \ C_w, \\
0, & \mathrm{on} \ C_g, \\
\frac{J_v(y,z)}{j \omega (\varepsilon_r-1) \varepsilon_0} , & \mathrm{in} \ C_v,
\end{dcases}
\end{align}
where $Z(y)$ is taking the impedance value $Z_n$ on the segment covered by the $n$-th impedance sheet and $\omega=2\pi f$ is the angular frequency. 

The system of integral equations in \eqref{eq:VSIE} can be solved with a Method of Moments (MoM).  To apply the MoM, the induced currents $J_w$, $J_g$ and $J_v$ are expanded to pulse basis functions that cover the respective domains. Specifically, the current density on each impedance sheet is segmented to $N_\mathrm{seg}$ 1-D pulse basis functions of width $\Delta_w=w_\mathrm{wire}/N_\mathrm{seg}$, totalling $N_w=N_\mathrm{seg}N$ functions for the whole impedance layer. The current density on the ground plane is segmented to $N_g$ 1-D pulse basis functions of width $\Delta_g=L_\mathrm{tot}/N_g$. Lastly, the volumetric polarization current in the substrate is segmented to $N_v=N_{v,y} \times N_{v,z}$ 2-D pulse basis functions with side lengths $\Delta_{v,y}=L_\mathrm{tot}/N_{v,y}$ and $\Delta_{v,z}=h/N_{v,z}$ along the $y$ and $z$ directions, respectively. By applying point-matching at the center of each cell, a linear-system of equations can be formed:
\begin{align} \label{eq:MoM}
\begin{pmatrix}
& \mathbf{G}_{gg} & \mathbf{G}_{gv} & \mathbf{G}_{gw} \\
& \mathbf{G}_{vg} & \mathbf{G}_{vv}-\mathbf{P} & \mathbf{G}_{vw} \\
& \mathbf{G}_{wg} & \mathbf{G}_{wv} & \mathbf{G}_{ww}-\mathbf{Z}_{w}
\end{pmatrix}
\begin{pmatrix}
\bar{J}_g \\
\bar{J}_v \\
\bar{J}_w
\end{pmatrix} =
\begin{pmatrix}
\bar{E}_g^\mathrm{inc} \\
\bar{E}_v^\mathrm{inc} \\
\bar{E}_w^\mathrm{inc}
\end{pmatrix}.
\end{align}
The vectors $\bar{E}_i^{inc}$ and $\bar{J}_i$ contain the sampled incident field and current values, respectively, at the center of each discretization cell on the impedance sheets ($i=w$), on the ground plane ($i=g$) and in the dielectric cross-section ($i=v$). The matrices $\mathbf{G}_{ij}$ represent the mutual and self interactions of the various cells and their analytic expressions can be found in the Appendix~\ref{app:A}. Finally, the polarization matrix $P$ is a diagonal matrix with all diagonal elements equal to $1/(j \omega (\varepsilon_r-1) \varepsilon_0)$ and $Z_w$ is a diagonal matrix with its elements taking the impedance values of the respective sheet segments.

Solving the system of Eq.~\eqref{eq:MoM} allows to calculate the induced currents for a particular set of impedances and geometric parameters at a specified frequency $f$. Subsequently, the reflected field $E^{ff}(\theta)$ in the far-field can be calculated as a summation of the far-field contributions from each discretized induced current $J_i[n]$. The directivity at each angle $D(\theta)$ is finally calculated based on standard definitions, which are also included in the Appendix~\ref{app:A}.

\subsection{Multi-frequency optimization}
While the analysis above was outlined for a single frequency, a set of frequencies can be examined simultaneously. The basis/testing functions involved in the Method of Moments are identical for all frequencies. However, the interactions matrices $\mathbf{G}_{ij}$, the sampled incident field $\bar{E}_i^{inc}$, the polarization matrix $P$ and impedance matrix $Z_w$ are all varying with frequency. With the exception of the impedance matrix, the other matrices are defined based on the geometric and material parameters and, therefore, they can be calculated only once. On the other hand, the imaginary part of each impedance sheet is optimized to achieve the desired response. The optimization refers to the imaginary part of the impedance in the center frequency $f_0$ and it is performed within a realizable range $X_n (f_0) \in [X_\mathrm{min}(f_0), X_\mathrm{max}(f_0)], \forall n$. The imaginary part of the impedance at another frequency $f$ is calculated based on a dispersion model that is provided a-priori and corresponds to the unit cell to be used in the final structure. In this way, it is guaranteed that the cell dispersion abides by the dispersion model at each examined frequency $f$. Finally, the real part $R_n(f)$, that corresponds to ohmic losses, is also supplemented at each frequency based on a lookup table containing this information for the unit cell.

Forming the system in Eq.~\eqref{eq:MoM} for multiple frequencies $f_m, m=1,..,N_f$, where $N_f$ is the number of sampling points within a frequency range, it is possible to analyze the structure at multiple frequencies simultaneously and calculate its directivity pattern $D(\theta,f_m)$. Furthermore, a multi-frequency cost function can be defined over all frequencies $f_m$. Focusing on the case of achromatic anomalous reflectors, the illumination efficiency at the desired reflected angle $\theta_r^0$ is first defined as:
\begin{align}\label{eq:e_il}
e_\mathrm{il} (f)= \frac{D(\theta_r^0,f)}{2 \pi (L_\mathrm{tot}/\lambda) \mathrm{cos}(\theta_r^0)},
\end{align}
where $\lambda = c_0/f$ is the free-space wavelength at the particular frequency. The illumination efficiency quantifies how the obtained directivity compares to the directivity of an aperture antenna with uniform amplitude and linear phasing to tilt the radiation towards the angle $\theta_r^0$. Aiming to the highest possible illumination efficiency over all frequency points, the cost function $F$ is then defined as the negative of the minimum illumination efficiency, namely:
\begin{align} \label{eq:costFunc}
F= - \min_m \{e_\mathrm{il}(f_m)\}, \ \ m=1,..,N_f.
\end{align}
Minimization of the cost function $F$ with respect to the impedance values $X_n(f_0)$ guarantees that a high directivity at the fixed angle $\theta_r^0$ will be obtained across all frequency points.

The optimization is performed in MATLAB with the use of the built-in functions \textit{ga} and \textit{fmincon} implementing a genetic algorithm and gradient descent optimization, respectively. A multi-stage optimization procedure is adopted to arrive to the final solution. In the first step, an initial point is obtained by minimizing a modified cost function $F=-\mathrm{mean} \{e_\mathrm{il}\}$, referring to the average efficiency over the selected frequencies points that are separated by $\Delta f_1$. This initial step consists of a genetic algorithm with $20$ generations and a total of approximately $4000$ objective function evaluations, followed by a gradient descent of a maximum of $500$ iterations. The genetic algorithm serves as a pseudo-random feed for the gradient descent that is much faster, considering that the gradient is semi-analytically calculated and provided to the solver at each iteration. In the second step, the previous optimized solution serves as a starting point, the frequency range is discretized further to $\Delta f_2$ and gradient-descent optimization is performed with the cost function of Eq.~\eqref{eq:costFunc}. The different definition at the two steps was selected after observations that the global optimization part performs better with the target of the average illumination efficiency over the frequency range, which is intuitively a smoother function compared to the efficiency of the worst frequency point.
Additionally, the multi-stage optimization reduces the computational time by allowing the global optimization to be performed over fewer frequency points, whereas the second part that relies on the faster gradient-descent method can be performed over more frequency points to avoid sharp dips of the illumination efficiency. 

\subsection{Unit cell and impedance characterization}
Each unit cell of the metasurface is represented with a rectangular strip of width $w$ and frequency-dispersive homogenized surface impedance $Z(f)$. Therefore, a way to match physical cells to impedance values and vice versa needs to be established. To this purpose, a full-wave simulation of a single unit cell is performed in Ansys HFSS. The unit cell is placed in the middle of the impedance layer, while still being on top of the grounded dielectric layer. The cell is illuminated by an embedded current line-source in the middle of the substrate ($z=h/2$,$x=0$), and the scattered field is recorded in a line segment $x\in [-\lambda_0,\lambda_0]$, $z=h+\lambda/2$ above the unit cell. The simulated scattered field is compared with the predicted scattered field from the MoM model when a single homogenized strip of varying $Z(f)=R(f)+jX(f)$ is illuminated by the same current line-source. The impedance value with the closest matching in terms of the scattered field is selected as the impedance value of the particular geometry. The process is repeated for a set of frequencies and for cells with different geometric parameters.

\begin{figure}[t]
\centering
\includegraphics[width=0.99\columnwidth]{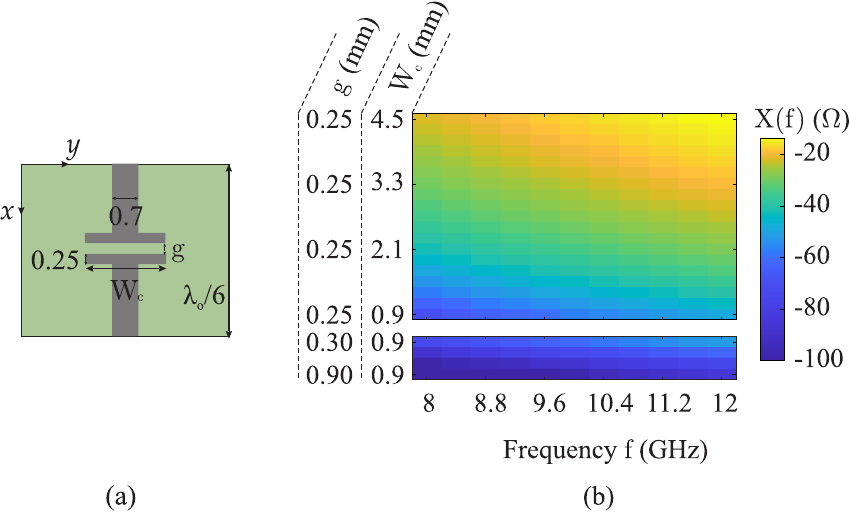}
\caption{\label{fig:Fig2} (a) Sketch of the unit cell geometry (dimensions in mm). The substrate has a thickness of $h=1.52 \ \mathrm{mm}$ and relative permittivity $\varepsilon_r=3$. (b) Imaginary part of the surface impedance $X(f)=\mathrm{Im}\{Z(f)\}$ for different pairs of the width $W_c$ and gap $g$ of the printed capacitor.}
\end{figure}

The unit cell used throughout this work is depicted in Fig.~\ref{fig:Fig2}(a), and it takes the very simple form of a wire loaded with printed capacitors every $\Lambda_x=\lambda_0/6 \approx 5 \ \mathrm{mm}$, where the wavelength $\lambda_0$ refers to the center frequency $f_0=10 \mathrm{GHz}$. A simple design was purposefully selected to demonstrate that there is no requirement to engineer the cell dispersion in order to arrive to an achromatic response from the whole metasurface. The width of the printed capacitor $W_c$ is varied from $0.9\mathrm{mm}$ to $4.5\mathrm{mm}$ with a fixed gap $g=0.25 \mathrm{mm}$, while more negative reactance values can be obtained by fixing $W_c=0.9 \mathrm{mm}$ and increasing the gap $g$. For an underlying substrate of relative permittivity $\varepsilon_r=3$ and thickness $h=1.52 \ \mathrm{mm}$, the imaginary part $X(f)$ of the extracted surface impedance in the range $[8,12] \mathrm{GHz} $ is given in Fig.~\ref{fig:Fig2}. On the other hand, the real part $R(f)$ takes low values ($<2 \Omega$ for all geometries and frequencies), and it has a negligible impact on the predicted directivity pattern.

\section{Simulated Results}
\label{sec:simulation}
\subsection{Achromatic reflection for different reflected angles and bandwidth}
The ability of the metasurface to efficiently act as an achromatic reflector is studied with respect to the defined frequency range and the reflected angle. For simplicity, we assume that the incident angle is kept fixed at broadside ($\theta=0^\circ$). Moreover, the center frequency is $f_0=10 \mathrm{GHz}$ and the metasurface consists of $N=56$ impedance sheets along a total length of $L_\mathrm{tot}=8\lambda_0$. The substrate is assumed to be Rogers RO3003 with relative permittivity $\varepsilon_r=3$ and thickness $h=1.52 \ \mathrm{mm}$. Three frequency ranges are examined, in particular $[9.4,10.6] \ \mathrm{GHz}$, $[8.8,11.2] \ \mathrm{GHz}$ and $[8.2,11.8] \ \mathrm{GHz}$, which correspond to $12 \%$, $24\%$ and $36\%$ of fractional bandwidth, respectively. For each frequency range, the optimization is performed for reflected angles varying from $-60^\circ$ to $-15^\circ$ with a step of $5^\circ$, whereas the symmetric cases (i.e., reflected angles from $+15^\circ$ to $+60^\circ$) can be realized by reversing the order of impedance values. 

The optimization of the design is based on the framework described in Sec.~\ref{sec:optimization}. In particular, the current densities are discretized into $N_g=400$ one-dimensional ($1$-D) pulse basis functions across the ground plane, $N_w=N \times N_\mathrm{seg}=56 \times 5=280$ $1$-D basis functions across the impedance sheets and $N_v=N_{v,z} \times N_{v,y}= 6 \times 400=2400$ two-dimensional (2-D) pulse basis functions in the dielectric region. The calculation of the far-field radiation is performed at steps of $0.5^\circ$ in the reflected half plane. For all cases, the frequency points are selected every $\Delta f_1=0.2 \ \mathrm{GHz}$  for the first step of the optimization and every $\Delta f_2=0.05 \ \mathrm{GHz}$ in the second step. The optimization scheme is performed multiple times with slightly different results due to the pseudo-randomness of the genetic algorithm. For each design case, $20$ iterations were performed in the first step (based on the optimization of average efficiency), and the best $10$ cases were pursued in the second-step (based on the minimum efficiency). The best optimized solution after the second step is recorded and analysed. The whole optimization procedure (including all iterations) to produce the solutions for a single design case takes an average of $86 \ \mathrm{min}$, $133 \ \mathrm{min}$ and $167 \ \mathrm{min}$ in a standard personal computer for a $12\%$, $24\%$ and $36\%$ bandwidth, respectively.

\begin{figure}
\centering
\includegraphics[width=0.80\columnwidth]{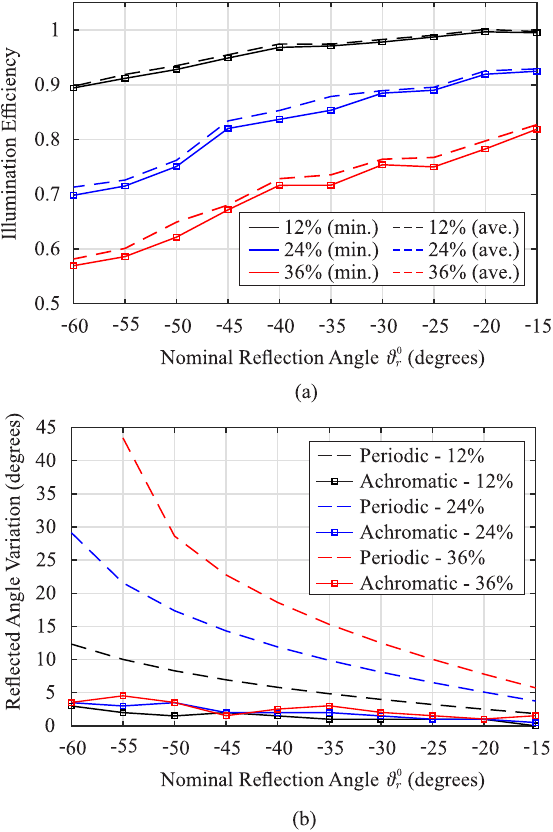}
\caption{\label{fig:Fig3} (a) Illumination efficiency (minimum and average) as a function of the reflected angle $\theta_r^0$ for designs illuminated from $\theta_i=0^\circ$ and operating over different bandwidth ranges centred at $f_0=10\mathrm{GHz}$. Metrics are calculated through the integral-equation framework for each optimized solution. (b) Maximum variation of the angle of maximum radiation $\theta_r$ within the selected fractional bandwidth ($12\%$, $24\%$, or $36\%$) for designs aiming at different desired reflected angles $\theta_r^0$. The optimized achromatic solutions (solid lines) are compared with the angle variation for a periodic or phase-gradient MTS over the same fractional bandwidth (dashed lines).}
\end{figure}

The minimum and average illumination efficiency over the examined bandwidth, as defined in Eq.~\eqref{eq:e_il}, is given for all reflected angles and bandwidths in Fig.~\ref{fig:Fig3}. Naturally, requiring the achromatic reflector to operate over a broader bandwidth reduces the illumination efficiency that can be achieved at a fixed angle over the whole frequency range. This clearly demonstrates a trade-off that can be utilized when designing passive reflectors that require an achromatic response over a specified bandwidth.

As mentioned above, a periodic structure would unavoidably suffer from chromatic aberrations due to the emergence of Floquet-Bloch modes. Assuming normal incidence, the periodicity of such a structure would be fixed to $\Lambda= m \lambda_0/|\mathrm{sin}(\theta_r^0)|$, so that the $m-$th mode reflects at an angle $\theta_r^0$ in the center frequency $f_0$. However, it can be shown that the reflected angle of that mode shifts with frequency as:
\begin{align}\label{eq:angle_var}
\theta_r (f)= \mathrm{sin}^{-1} \left[\frac{f_0}{f} \mathrm{sin} (\theta_r^0)\right].
\end{align}
Notably, the same result is obtained for a phase-gradient reflective metasurface designed through the generalized law of reflection at the center frequency $f_0$, if the reflection coefficient is assumed to be locally dispersionless. The angle variation in Eq.~\eqref{eq:angle_var} is more significant for a broader bandwidth and a wider reflected angle. Intuitively, this explains why achieving high illumination efficiency for such cases is more challenging, as noticed in Fig.~\ref{fig:Fig3}(a). For the three bandwidth ranges and choices of output angles investigated herein, the angle variation of a periodic (or phase-gradient) metasurface is plotted in Fig.~\ref{fig:Fig3}(b). In comparison, the angle variation for each one of our optimized solutions are also given (with a $0.5^\circ$ quantization) by analyzing the optimized impedance metasurfaces through the integral-equation framework. Arguably, the optimized designs exhibit a much tighter range for the reflected angle over the examined bandwidth. Specifically, the angle of maximum radiation varies less than $3^\circ$, $3.5^\circ$ and $4.5^\circ$ for a bandwidth of $12\%$, $24\%$ and $36\%$, respectively, and for a reflected angle up to $\pm 60^\circ$. It is also noted that the angle variation was not explicitly included in the cost function as an additive term or as a constraint. On the contrary, aiming for high directivity at the desired output angle $\theta_r^0$, naturally brings the angle of maximum radiation $\theta_r$ close to that angle.

\subsection{Role of auxiliary surface waves}
It is worth examining the underlying mechanism enabling the passive anomalous reflection at a fixed angle over a wide bandwidth. The presented MTS features a distance of $\lambda_0/7$ between consecutive cells, which is considerably denser than traditional reflectarray implementations having elements typically at the size of $\lambda_0/2$. In turn, this dense array of unit cells allows the excitation of surface waves that manifest as high spectral components of the scattered electric field (in terms of a $k_y$ wavenumber), or in terms of high spectral components of the induced currents. These evanescent fields (also, referred to as non-radiating currents) have been utilized in previous works to render the impedances passive and lossless, by restoring the local power conservation condition and redistributing the incident power along the MTS boundary \cite{Epstein:PRL2016,Kwon:PRB2018,Kwon:AWPL2021,Ataloglou:AWPL2021,Salucci:TAP2018}. Importantly, the integral-equation framework adopted in this work allows to accurately capture all near-field interactions compared to approaches that adopt a cell-to-cell design and suffer from unaccounted coupling \cite{Xu:TAP2019}.

\begin{figure}
\centering
\includegraphics[width=0.65\columnwidth]{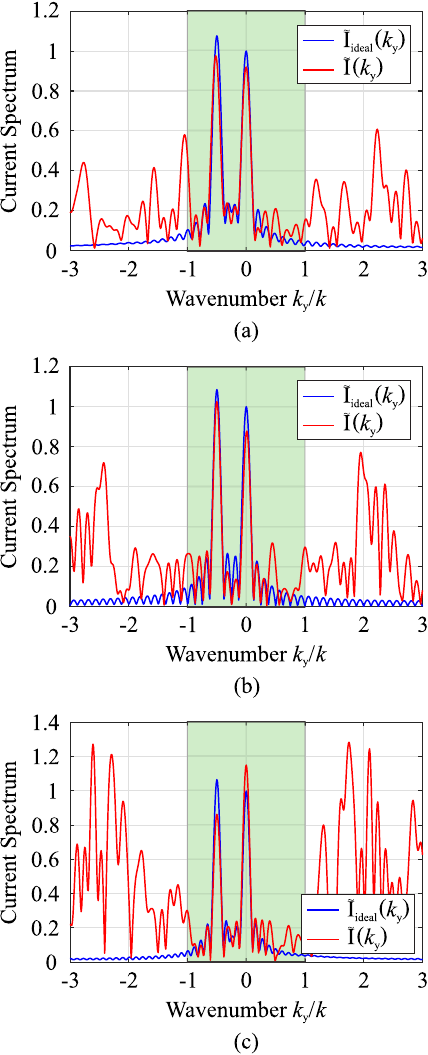}
\caption{Spectrum $\tilde{I}(k_y)$ of the induced currents on the optimized impedance sheets for the design case with reflected angle $\theta_r^0=-30^\circ$ and a bandwidth of $\mathrm{BW}=24\%$. The spectrum $\tilde{I}_\mathrm{ideal} (k_y)$ of a current array that fully suppresses the specular radiation and radiates a beam in the desired reflected direction is also included for comparison. The normalized spectrum is plotted for (a) $f=8.8 \ \mathrm{GHz}$, (b) $f=10 \ \mathrm{GHz}$ and (c) $f=11.2 \ \mathrm{GHz}$ \label{fig:Fig7}}
\end{figure}

To verify the emergence of evanescent fields in the optimized solutions, we investigate the spectrum of the excited currents for the design exhibiting $24\%$ bandwidth and a reflected angle of $\theta_r^0=-30^\circ$ for broadside incidence ($\theta_i=0^\circ$). The induced current over each impedance sheet $I[n]$ is calculated through the integral-equation framework for every frequency. A spatial Fourier transform is then performed on the discrete signal as:
\begin{align}
\tilde{I}(k_y)=\sum_{n=0}^N I[n] \mathrm{exp}\{+j k_y y_c[n]\},
\end{align}
where $y_c[n]$ refers to the coordinate of the center of the $n$-th cell. We also calculate the spectrum $\tilde{I}_\mathrm{ideal} (k_y)$ of an induced current array $I_\mathrm{ideal}[n]$ that consists of two components, namely one component that cancels the specular reflection and one that produces a beam at the desired angle $\theta_r^0$ with a total scattered power equal to the total incident power. Both spectrums are normalized to $\tilde{I}_\mathrm{ideal} (k_y=0)$ and they are plotted for the center and edge frequencies in Fig.~\ref{fig:Fig7}. As noted, two peaks can be observed for the spectrum of the induced currents, located in the specular direction ($k_y=0$) and in the desired reflected direction ($k_y/k=\mathrm{sin}(\theta_r^0)=0.5$). The propagating components ($|k_y|<k_o$) remain, in general, close to the ideal case, with any differences manifesting in the radiation pattern as perturbed minor lobes or slightly lower efficiency in the main lobe with regards to the ideal case. However, the spectrum for the optimized induced currents shows significant evanescent components, especially for the frequencies in the higher end of the operational bandwidth. These auxiliary evanescent components serve a dual purpose. First, they render the homogenized impedances passive and almost lossless (i.e., having only the small losses $R(f)$ as dictated by the simulation of the actual cell). This has been established through many works that have shown that just stipulating the propagating fields or currents leads to complex impedances, designating lossy or active regions across the MTS. In addition to restoring passivity, the excited frequency-dependent evanescent components in this work make the homogenized impedances $Z_n(f)$ compatible with the dispersion model $Z(f)$ corresponding to the actual patterned cell.

\section{Experimental Verification}
\label{sec:measurement}
\subsection{Fabricated Prototypes and Measurement Setup}
\begin{figure}
\centering
\includegraphics[width=0.99\columnwidth]{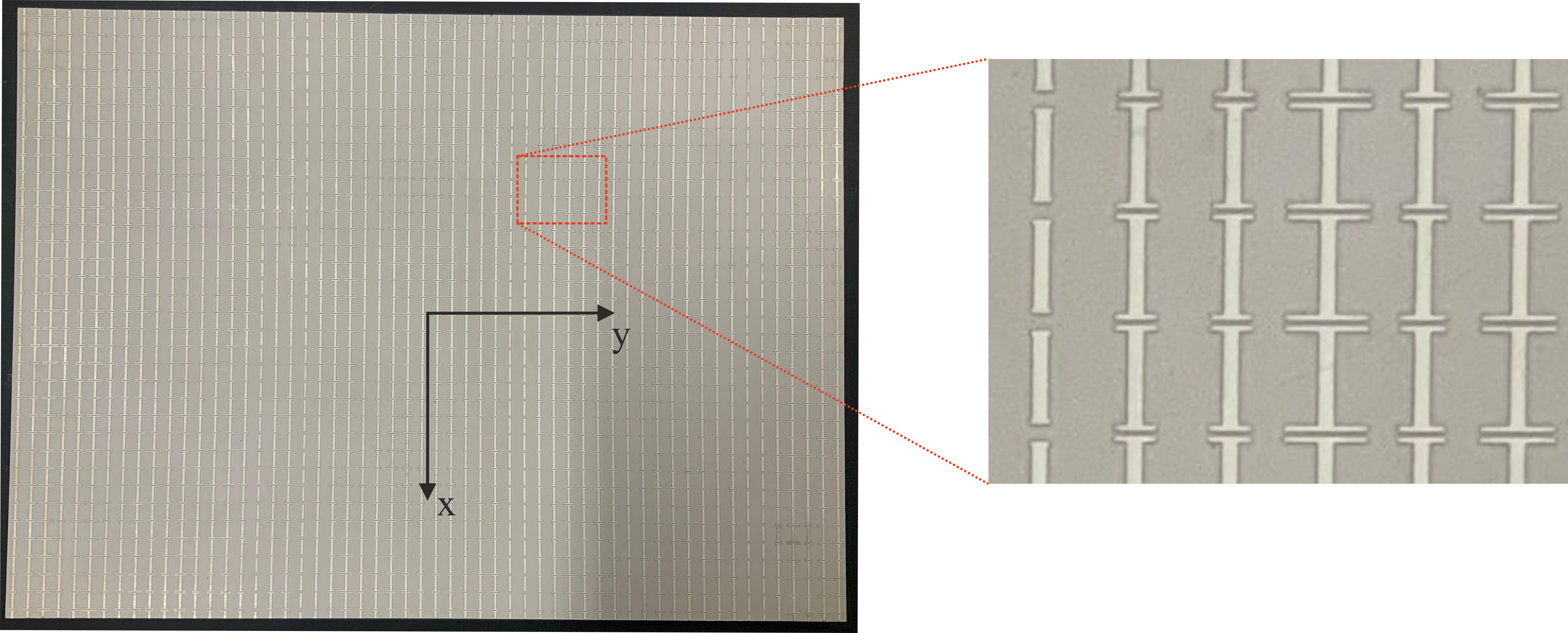}
\caption{\label{fig:Fig4} Photograph of the fabricated prototype for reflection at $\theta_r^0=-30^\circ$ over $24\%$ bandwidth.}
\end{figure}

To further verify our framework, three achromatic anomalous reflectors are fabricated and measured. Similar to the analysis in the previous section, all reflectors are illuminated by a normally-incident plane wave ($\theta_i=0^\circ$) and different reflected angles and bandwidth around $f_0=10\mathrm{GHz}$ are desired. In particular, the designs aim to achieve anomalous reflection to $\theta_r^0=30^\circ$ over a $24\%$ and a $36\%$ bandwidth, and anomalous reflection to $\theta_r^0=45^\circ$ over a $24\%$ bandwidth. The designs are first simulated with Ansys HFSS, using the proposed capacitively-loaded cells in Fig.~\ref{fig:Fig2}(a) to realize the optimized homogenized impedances. It is noted that the substrate has a relative permittivity of $\varepsilon_r=3$ and a thickness of $h=1.52 \ \mathrm{mm}$, which corresponds to approximately $0.05 \lambda_0$ at the center frequency. The fabricated boards are truncated to $\approx 5.83 \lambda_0$ along the $x$-axis, which was a sufficient size in simulations to achieve an almost identical response compared to the infinitely-long case. A photograph of the fabricated prototype for the first design ($\theta_r^0=-30^\circ$, $24\%$ bandwidth) is given in Fig.~\ref{fig:Fig4}.

\begin{figure}[t]
\centering
\includegraphics[width=0.85\columnwidth]{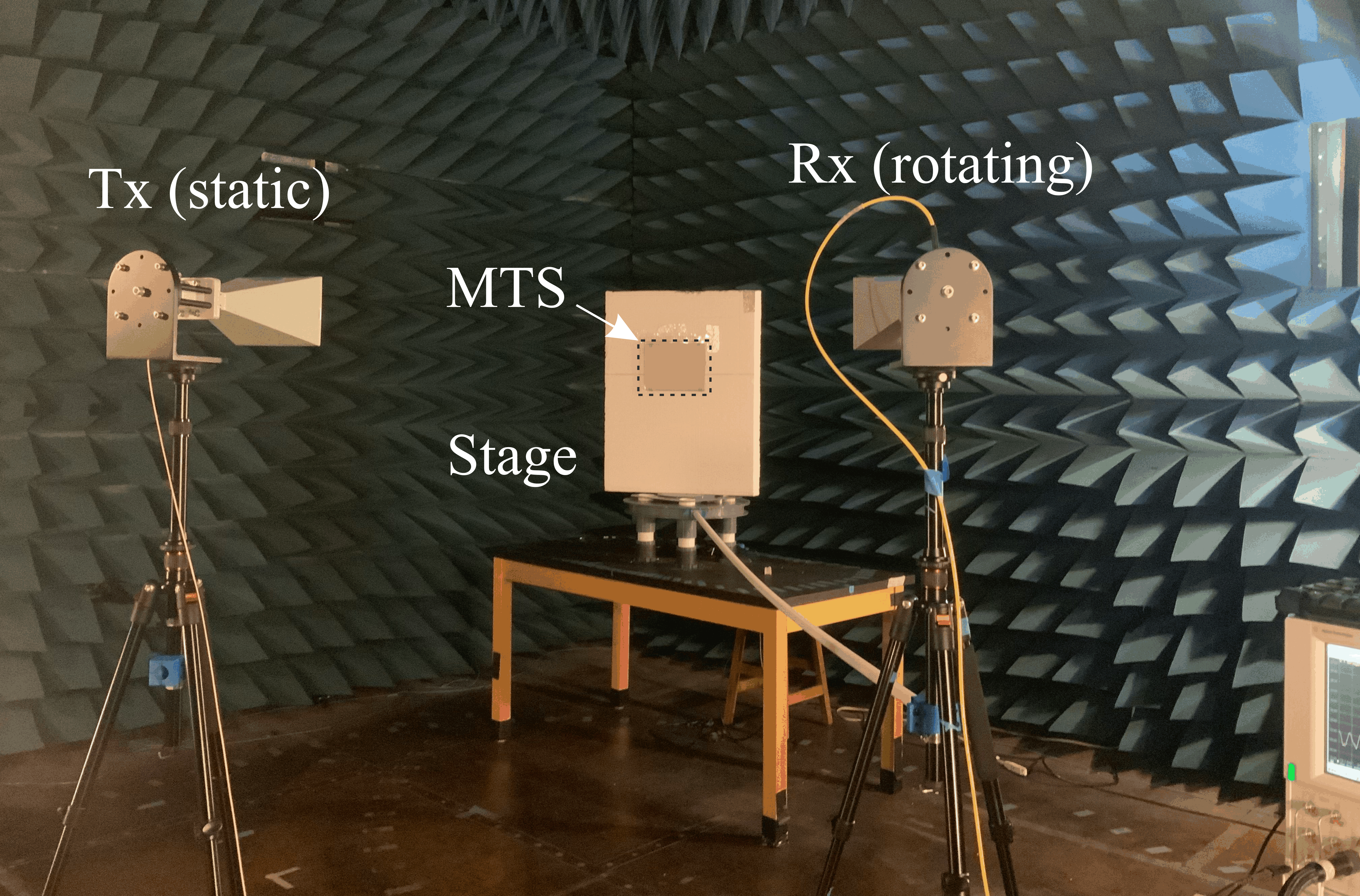}
\caption{\label{fig:setup} Photograph of the bistatic measurement setup.}
\end{figure}

A bistatic measurement setup is utilized for the measurement of the reflected field in the $yz$-plane, as shown in Fig.~\ref{fig:setup}. Specifically, a horn antenna illuminates the metasurface from the broadside direction and a distance of $r_1=2.42 \mathrm{m}$ (between the aperture of the antenna and the center of the MTS), while a second horn antenna is rotated at a constant radius $r_2=2.45 \mathrm{m}$ in order to capture the reflected field. Measurements are recorded with a Vector Network Analyser (VNA) every $1^\circ$ within a range of $\pm 20^\circ$ around the desired reflected angle and every $2.5^\circ$ for the rest of the pattern. Due to the size of the antennas, a zone around retro-reflection is formed that $S_{21}$ cannot be measured. To minimize this zone, the reflected angles of $\pm 5^\circ$ and $\pm 2.5^\circ$ are measured by utilizing the diagonal of the chamber and moving the antennas to distances $r_1'=5.55 \mathrm{m}$ and $r_2'=5.56 \mathrm{m}$. The $S_{21}$ measurements for these reflected angles are then scaled through the bistatic radar equation to the expected measurements at the common shorter distances.

\subsection{Measured Results}
\begin{figure}
\centering
\includegraphics[width=0.65\columnwidth]{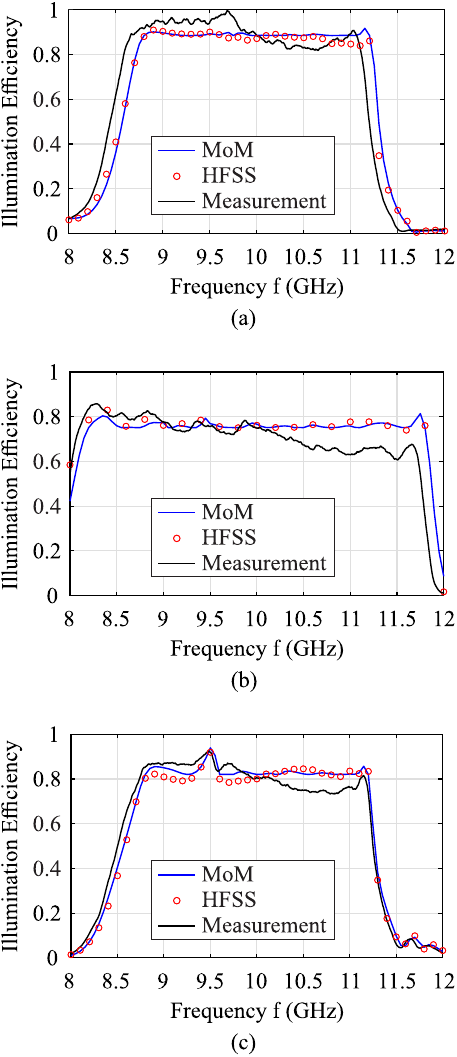}
\caption{Illumination efficiency for the achromatic reflectors illuminated from $\theta_i=0^\circ$ and aiming a reflected angle: (i) $\theta_r^0=-30^\circ$ over $24\%$ bandwidth, (ii) $\theta_r^0=-30^\circ$ over $36\%$ bandwidth, and (iii) $\theta_r^0=-45^\circ$ over $24\%$ bandwidth, respectively. The MoM solution is compared with HFSS simulations of the realistic structure and with the measured results. \label{fig:Illumination_efficiency}}
\end{figure}

For the illumination efficiency, the $2$-D directivity is calculated at each frequency point from the measured $|S_\mathrm{21} (\theta)|$ values (in linear scale) as:
\begin{align}\label{eq:D_meas}
D(\theta)= 2\pi \frac{|S_{21}(\theta)|^2}{\int_{-\pi/2}^{\pi/2} |S_{21} (\theta)|^2 d\theta},
\end{align}
where the integration is performed numerically over the recorded angles. The illumination efficiency $e_\mathrm{il}$ is then defined at the desired angle $\theta_r^0$ for each design case, as in Eq.~(4). The illumination efficiency as a function of frequency is given in Fig.~\ref{fig:Illumination_efficiency}(a)-(c) for the MoM solution of the integral-equation framework, for the HFSS full-wave simulations, and for the measured results. For all designs, the measured results match quite well with the predictions over the desired bandwidth. Specifically, the first design ($\theta_r=-30^\circ$, $\mathrm{BW}=24\%$) shows a minimum illumination efficiency of $81.8\%$ over the range $[8.7,11.1] \mathrm{GHz}$ with the average value reaching $90.1\%$. Similarly, the second design ($\theta_r=-30^\circ$, $\mathrm{BW}=36\%$) demonstrates a minimum efficiency of $61.1\%$ (average $73.1\%$) for the frequencies $[8.1,11.7] \mathrm{GHz}$. The small offset of $0.1\ \mathrm{GHz}$ in the frequency band compared to the nominal range can be attributed to a slightly increased dielectric constant for the substrate. Comparing the two designs, it is confirmed that an achromatic anomalous reflector can be designed over a higher bandwidth at the expense of a less directive beam. Lastly, the third design achieves reflection at a wider angle ($\theta_r=-45^\circ$, $\mathrm{BW}=24\%$) over the frequency range $[8.8,11.2] \mathrm{GHz}$ with a minimum illumination efficiency of $73.3\%$ (average $81.4\%$). It is noted that the illumination efficiency takes into account the directivity at the fixed reflected angle $\theta_r^0$ that is desired for each design; thus, small variations of the angle of maximum radiation impact negatively the metric on top of any broadening of the main lobe.

\begin{figure}
\centering
\includegraphics[width=0.65\columnwidth]{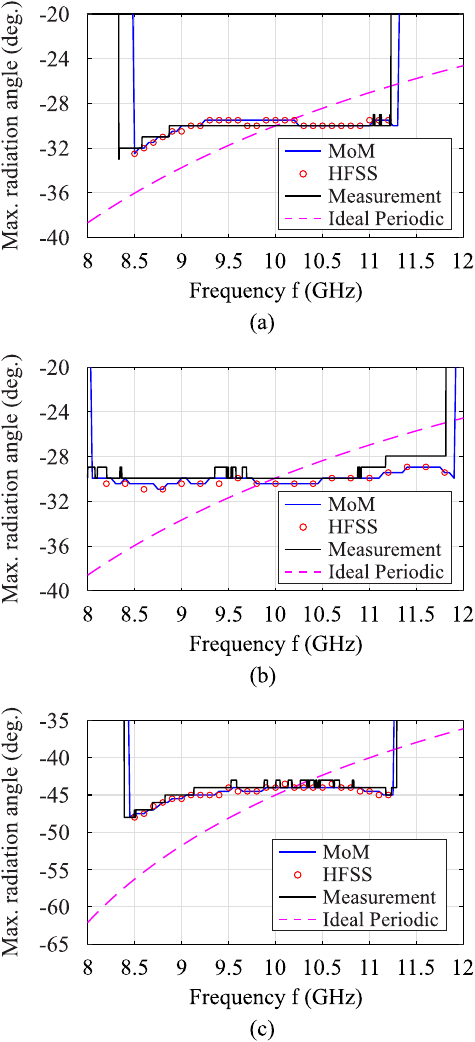}
\caption{The angle of maximum radiation $\theta_r$ as a function of frequency for the designs: (i) $\theta_r^0=-30^\circ$ over $24\%$ bandwidth, (ii) $\theta_r^0=-30^\circ$ over $36\%$ bandwidth, and (iii) $\theta_r^0=-45^\circ$ over $24\%$ bandwidth, respectively. The angle variation for the achromatic reflectors is compared with that of a periodic structure designed at $f_0=10 \ \mathrm{GHz}$. \label{fig:Angle_variation}}
\end{figure}

\begin{figure}
\centering
\includegraphics[width=0.65\columnwidth]{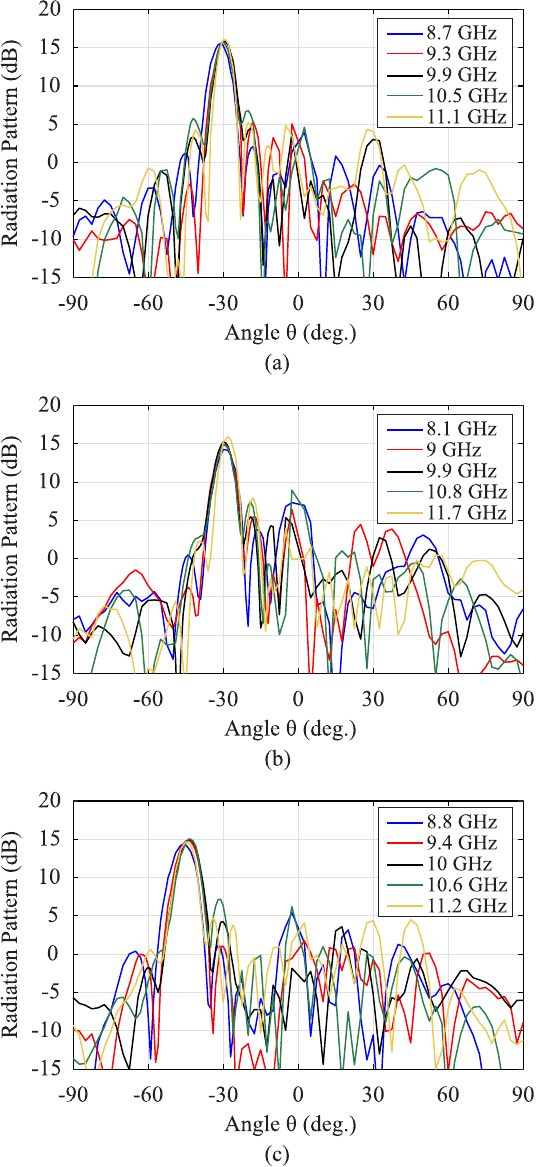}
\caption{Measured radiation patterns for $5$ individual frequencies across the observed bandwidth for the designs aiming at: (a) $\theta_r^0=-30^\circ$, $\mathrm{BW}=24\%$, (b) $\theta_r^0=-30^\circ$, $\mathrm{BW}=36\%$, and (c) $\theta_r^0=-45^\circ$, $\mathrm{BW}=24\%$. \label{fig:Fig6}}
\end{figure}

The reflected angle $\theta_r$ where the maximum radiation is observed is plotted for the $3$ cases in Fig.~\ref{fig:Angle_variation}(a)-(c). Across the operational frequency ranges reported above, the measured angle variation is $1.5^\circ$, $2^\circ$ and $3^\circ$, respectively, showing good agreement with the predicted angle variations from the MoM or simulation analyses. The angle variation for a periodic (or a phase-gradient) MTS, as calculated in Eq.~\eqref{eq:angle_var} for a design at the center frequency $f_0=10 \mathrm{GHz}$, is also plotted for comparison. As shown, the achromatic anomalous reflectors considerably reduce the angular variation by enforcing a high directivity across a desired frequency range. The angular stability and high directivity of the beam can also be verified by the measured directivity patterns in different frequencies. In particular, $5$ equally-spaced frequencies across the observed bandwidths are plotted for all three cases in Fig.~\ref{fig:Fig6}. It should also be mentioned that the selected bandwidth of operation for all three designs far exceeds the calculated upper bound for achromatic reflection under the assumption of single-resonance cells \cite{Fathnan:IoP2020}, that would estimate a $7\%$ and $5\%$ fractional bandwidth for a reflected angle $\theta_r^0=-30^\circ$ and $\theta_r^0=-45^\circ$, respectively (assuming $f_\mathrm{min}=8.8\mathrm{GHz}$).

\begin{figure}
\centering
\includegraphics[width=0.65\columnwidth]{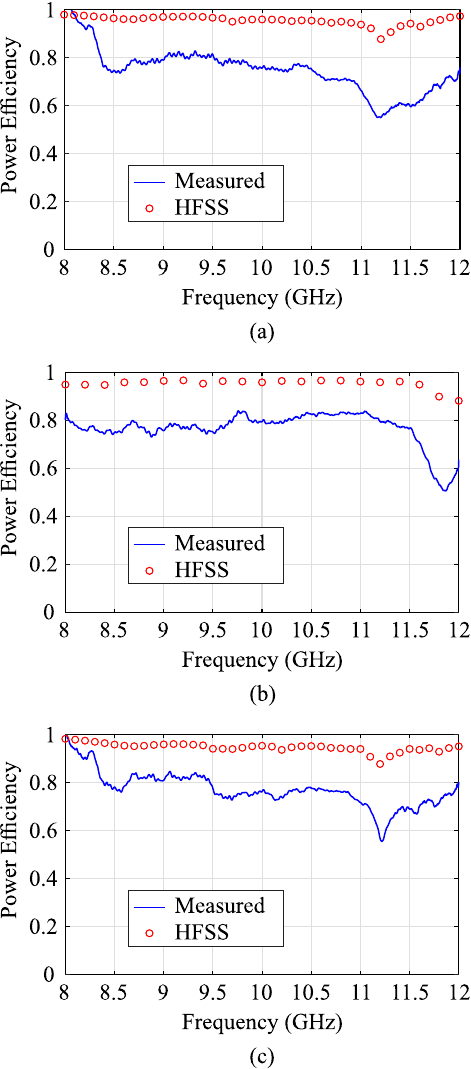}
\caption{Estimated power efficiency $e_p (f)$ for the three fabricated prototypes: (a) $\theta_r^0=-30^\circ$ over $24\%$ bandwidth, (b) $\theta_r^0=-30^\circ$ over $36\%$ bandwidth, and (c) $\theta_r^0=-45^\circ$ over $24\%$ bandwidth. The power efficiency of the HFSS simulated designs is also included.\label{fig:power_efficiency}}
\end{figure}

Finally, the power efficiency defined as the total reflected power divided by the incident power can be estimated. For this purpose, a total efficiency is first calculated as the ratio between the measured received power at the desired angle $\theta_r^0$ and the expected received power from an ideal anomalous reflector. Then, the total efficiency is split into the drop due to broadening in the azimuthal plane and the drop due to power losses. The analytic expressions for this approximate calculation are given in Appendix~\ref{app:B}, whereas the simulated and measured power efficiency for all three cases as a function of frequency is plotted in Fig.~\ref{fig:power_efficiency}. As observed, the power efficiency drops at the higher end of the operating bandwidth for each design. This is attributed to the higher evanescent spectrum that is excited and contributes to ohmic losses due to the higher field values at the vicinity of the metasurface. The average measured power efficiency (across the operational bandwidth) is estimated to be $76\%$, $78.4\%$ and $76.9\%$ for the three prototypes. Although power losses appear to be higher when compared with simulations, a slight overestimation may be possible due to the applied method, as all factors affecting the received power (apart from the illumination efficiency in the $yz$ plane) are perceived as power losses. These factors include a small deviation of the incident field from an ideal plane-wave, the edge effects of the MTS that lead to a slight non-uniformity along the $x$-axis, a small warpage of the MTS prototypes and any minor alignment issues in the measurement setup.

\subsection{Comparison with existing literature}
It is useful to compare the performance of the achromatic reflectors designed with the proposed integral-equation framework with previous works at microwaves. In \cite{Yang:APL2016}, an achromatic reflector was realized at $f_0=11.2 \ \mathrm{GHz}$ based on dispersion engineering of the individual unit cells. The design achieved a $9\%$ fractional bandwidth with a reflection gain maintaining $70\%$ of its maximum value. However, the thickness of the MTS is $0.56 \lambda_0$ owing to unit cell type. Similarly, a refractive design was designed in \cite{Ji:LPR2022} for refraction at a relatively narrow angle of $20^\circ$ over a $24\%$ bandwidth centered around $11 \ \mathrm{GHz}$. The strict requirement for controlling both the reflected phase and dispersion of each cell necessitated multiple copper layers and a thickness of $0.81 \lambda_0$. In \cite{Zhu:OE2020}, the possibility to reduce the MTS structure to a single layer was examined through machine learning algorithms operating at a unit cell level to achieve the desired dispersive response. A $10\%$ bandwidth for a reflected angle of $22^\circ$ was reported with the average simulated reflected amplitude (given in terms of a radar cross-section of $\approx 14.5 \ \mathrm{dBm^2}$) revealing a total efficiency of $\approx 66 \%$ compared to the radar cross-section of an ideal reflector of the same-size. A design for a bigger bandwidth of $35\%$ for retro-reflection at $20^\circ$ was attempted in \cite{Jia:JoP2018} by partitioning the aperture to segments designed for different discrete frequencies. Due to the segmentation of the MTS aperture, the simulated radar cross-section of the total structure ($\approx 12-17 \ \mathrm{dBm^2}$) is considerably lower compared to the radar cross-section of the individual impedance modulations over the same aperture size ($\approx 20-27 \ \mathrm{dBm^2}$) or an ideal reflector of the same size ($26.5-29.6 \ \mathrm{dBm^2}$). Lastly, broadband periodic structures, such as the one presented in \cite{Qi:OE2022}, may achieve high reflection ($>80\%$) to the desired Floquet-Bloch mode over a wide bandwidth ($54\%$), but the reflected angle varies by $35^\circ$ within the examined frequency range, as observed in the presented results and predicted by Eq.~\ref{eq:angle_var}. In contrast, the designs presented herein combine the low-profile thickness ($\approx 0.05 \lambda_0$ at the center frequency) together with the ability to have relatively wide bandwidth and achromatic reflection at wider angles. Furthermore, the high illumination efficiencies show a high utilization of the whole aperture over each individual frequency, and the average power losses are maintained to around $1-1.2 \ \mathrm{dB}$.

\section{Conclusion}\label{sec:conclusion}
We have introduced a novel design method for realizing achromatic anomalous reflection with a low-profile planar metasurface. The proposed method relies on modeling the unit cells as impedance sheets with a frequency-dispersive homogenized surface impedance. This computationally-efficient model can be utilized to predict the response of the metasurface through an integral-equation framework, solved with a Method of Moments. Additionally, optimization can be performed to find the impedance values that achieve the desired achromatic functionality. Importantly, the adopted unit cell is very simple and there is not any effort to engineer the dispersion of individual unit cells, as in traditional achromatic designs. Instead, the optimization framework harnesses near-field interactions and the excitation of surface waves in order to reflect the beam towards a fixed desired angle over a wide bandwidth. 

Compared to achromatic anomalous reflectors at microwaves that have been reported in the literature, the proposed optimization method results in low-profile structures that demonstrate improved performance in terms of the directivity of the reflected beams, as well as the range of reflected angles or fractional bandwidth that can be realized. An analysis of the illumination efficiency for a variety of reflection angles up to $60^\circ$ and fractional bandwidth up to $36\%$ was presented. Three particular design cases were fabricated and the  measurements demonstrated achromatic anomalous reflection at $\theta_r^0=-30^\circ$ with $90.1\%$ and $73.1\%$ average illumination efficiency over a $24\%$ or $36\%$ fractional bandwidth (around $10 \ \mathrm{GHz}$), respectively, and achromatic anomalous reflection at $\theta_r^0=-45^\circ$ with a $81.4\%$ average illumination efficiency over a $24\%$ fractional bandwidth. The proposed framework can also be explored for the design of multi-layer reflective designs, as a way to increase the available degrees of freedom in the optimization and achieve even higher performance compared to the metrics presented for single-layer metasurfaces in Fig.~\ref{fig:Fig3}. Furthermore, while the focus of this work is the design of achromatic reflectors, the method can be expanded to transmissive metasurfaces that operate as achromatic lenses or refractors.

\appendices 
\section{Analytical expressions for the MoM solution} \label{app:A}
The elements of the block matrices $\mathbf{G}_{ij}$ in Eq.~\eqref{eq:MoM} are given as \cite{Xu:Access2022}:
\begin{subequations}\label{eq:G_mom}
\begin{itemize}[leftmargin=10pt]
\item for $n \neq m \ \mathrm{or} \ i \neq j, j \in \{g,w\} $,
\begin{align}
\mathbf{G}_{ij}[n][m]= -\frac{k \eta \Delta_j}{4} H_0^{(2)} (k r_{nm}),
\end{align}
\item for $n \neq m \ \mathrm{or} \ i \neq j, j = v $,
\begin{align}
\mathbf{G}_{ij}[n][m]= -\frac{\eta \pi r_o}{2} J_1(k r_o) H_0^{(2)} (k r_{nm}),
\end{align}
\item for $n=m, i=j, j \in \{g,w\} $,
\begin{align}
\mathbf{G}_{ij}[n][m]= -\frac{k \eta \Delta_j}{4} \left[1-j\frac{2}{\pi} \mathrm{log}\left(\frac{1.781 k \Delta_j} {4e} \right) \right],
\end{align}
\item for $n=m, i=j, j =v $,
\begin{align}
\mathbf{G}_{ij}[n][m]= -\frac{\eta }{2k} \left[kr_o H_1^{(2)} (kr_o) -2j \right],
\end{align}
\end{itemize}
\end{subequations}
where $r_{nm}$ is the distance between each source location to each observation point (both centered in the middle of the respective MoM cells), $J_1$ is the Bessel function of first order, $e \approx 2.718$ is Napier's constant and $r_0=\sqrt{\Delta_{v,y} \Delta_{v,z}/\pi}$. It is noted that for the matrices referring to the volumetric surface current density ($j=v$) the approach of approximating the cell area with a circular cross-section of radius $r_0$ is adopted \cite{Richmond:TAP1965}. The last two cases in Eq.~\eqref{eq:G_mom} refer to the self-interactions, i.e. when the testing and and basis functions are identical and $r_\mathrm{mn}=0$. For these cases, a small-argument approximation is considered for the Hankel function resulting in the listed expressions. 

Having the induced currents allows to calculate the scattered electric-field in the far field at a set of $N_{ff}$ angles distributed uniformly in the $\theta \in [-\pi/2, \pi/2]$ range. This is performed through a matrix multiplication:
\begin{align} \label{eq:Eff}
\bar{E}^{ff}= \begin{bmatrix} \mathbf{G}_{fg} &  \mathbf{G}_{fv} & \mathbf{G}_{fw} \end{bmatrix} \begin{bmatrix}
\bar{J}_g \\
\bar{J}_v \\
\bar{J}_w
\end{bmatrix}, 
\end{align}
where $\bar{E}^{ff}$ is a vector containing the electric-field value in the discretized angles $\theta_n$ ($n=1,..,N_{ff}$) and $\mathbf{G}_{fg}, \mathbf{G}_{fv}, \mathbf{G}_{fw}$ are matrices with respective sizes $N_{ff} \times N_g$,  $N_{ff} \times N_v$ and $N_{ff} \times N_w$, and expressions given for the individual elements $\mathbf{G}_{fi}[n][m]$ as:
\begin{subequations}
\begin{itemize}[leftmargin=10pt]
\item for $i=\{g,w\}$,
\begin{align} \label{eq:Gf_coefficients}
\hspace{-10pt} \mathbf{G}_{fi}[n][m]= -\frac{\eta \Delta_i}{4} \sqrt{\frac{2jk}{\pi}} \mathrm{exp} \{jk(y_m \mathrm{sin} (\theta_n) + z_m \mathrm{cos} (\theta_n))\},
\end{align}
\item for $i=v$,
\begin{align}
\hspace{-10pt} \mathbf{G}_{fi}[n][m]=-\frac{\eta \pi r_0^2}{4} \sqrt{\frac{2jk}{\pi}} \mathrm{exp} \{jk(y_m \mathrm{sin} (\theta_n) + z_m \mathrm{cos} (\theta_n))\},
\end{align}
\end{itemize}
\end{subequations}
where the coordinates $(y_m,z_m)$ refer to the center of the respective current basis function. It is evident by the calculation in Eq.~\eqref{eq:Eff}-\eqref{eq:Gf_coefficients} that the $1/\sqrt{r}$ decay of the electric field values $\bar{E}^{ff}$ is already suppressed. Therefore, the directivity is calculated for different angles $\theta_n$ as:
\begin{align} 
D(\theta_n)= 2\pi \frac{|E^{ff}(\theta_n)|^2}{\int_{-\pi}^{\pi} |E^{ff}(\theta)|^2 d\theta},
\end{align}
where the integration in the denominator is carried out numerically using the discrete values $E^{ff} (\theta_n)$. 

Lastly, it should be mentioned that the linear system in Eq.~(3) can be simplified through the Kron's reduction method \cite{Xu:Access2022}. Essentially, the two first rows of the system are utilized to express the sampled current densities $\bar{J}_g$ and $\bar{J}_v$ as a function of the current density $\bar{J}_w$. The third row can then be used to form a linear system only in terms of the sampled current density $\bar{J}_w$. Importantly, the modified linear system is of size $N_w \times N_w$ greatly reducing the computational cost for calculating the induced currents and far-field radiation at each algorithm iteration.

\section{Estimation of the power efficiency} \label{app:B}
The bistatic radar equation dictates that the square of the transmission coefficient $|S_{21}|^2$ through a reflection from a target of a bistatic cross-section $\sigma (\theta_r,\theta_i)$ would be \cite{Balanis_book}:
\begin{align} \label{eq:bistatic}
|S_{21} (\theta_r,\theta_i)|^2=\frac{P_r}{P_i}=\sigma(\theta_r,\theta_i) \frac{G_{t}G_{r}}{4\pi} \left(\frac{\lambda}{4\pi R_1 R_2}\right)^2,
\end{align}
where $G_t$ and $G_r$ are the gains of the transmitting and receiving antennas, respectively, and $R_1$, $R_2$ are the distances between each antenna and the target. The phase center of the antenna is considered as its position in all calculations. Thus, a small offset $d_c$ (ranging from $d_c=0.071\mathrm{m}$ at $8 \ \mathrm{GHz}$ to $d_c=0.148\mathrm{m}$ at $12 \ \mathrm{GHz}$) is added to the distance, as measured between the target and the antennas' apertures \cite{Muehldorf:TAP1970}. The antenna gain for the two identical horn antennas is experimentally calculated through the Friis Equation by aligning the two horn antennas to face each other with a distance of $r_{12}=2.40\mathrm{m}$ between the two apertures. The extracted gain values, assuming equal gain for the two identical horn antennas, match well with the theoretical values of the antenna's datasheet \cite{A-info}, with the maximum offset being $0.16 \ \mathrm{dB}$. 

An ideal reflector would reflect all the incident power from an incident angle $\theta_i$ to the desired direction $\theta_r^0$ with the directivity of a uniform-amplitude aperture that radiates towards that direction. For such an ideal, lossless structure with dimensions $L_x \times L_y$ the cross section would be:
\begin{align} \label{eq:sigma_ideal}
\sigma_\mathrm{ideal}(\theta_r^0,\theta_i)=4 \frac{(L_x L_y)^2}{\lambda^2} \mathrm{cos}(\theta_i) \mathrm{cos}(\theta_r^0).
\end{align}
Replacing the ideal (maximum) cross section of Eq.~\eqref{eq:sigma_ideal} to Eq.~\eqref{eq:bistatic} leads to the $|S_{21}^\mathrm{ideal} (\theta_r^0)|$ coefficient that would have been measured for an ideal anomalous reflector with the presented setup. The ratio between the measured received power from the prototype to the expected power from an ideal reflector, or equivalently, the measured cross-section of the MTS to the ideal cross-section, defines a total efficiency:
\begin{align}
e_\mathrm{tot}= \frac{|S_{21}(\theta_r^0)|^2}{|S_{21}^\mathrm{ideal}(\theta_r^0)|^2}=\frac{\sigma(\theta_r^0)}{\sigma_\mathrm{ideal}(\theta_r^0)}.
\end{align}
The total efficiency can be analyzed into two parts ($e_\mathrm{tot}=e_\mathrm{il} e_p$) with: (i) the illumination efficiency $e_\mathrm{il}$ defined as the obtained directivity divided by the directivity of a uniform-amplitude aperture, and (ii) the power efficiency  $e_p$ defined as the total reflected power divided by the total incident power. The illumination efficiency $\eta_\mathrm{il}$ is calculated first through Eq.~\eqref{eq:D_meas} applied in the $yz$-plane. Then, the power efficiency is computed at each individual frequency as:
\begin{align}
e_p=\frac{e_\mathrm{tot}}{e_\mathrm{il}}.
\end{align}
The extracted values for the three designs are plotted in Fig.~\ref{fig:power_efficiency}, along with the power efficiency calculated from full-wave simulations. 

%


\end{document}